\documentclass[runningheads]{cl2emult}

\usepackage{makeidx}  
\usepackage{graphicx} 
\usepackage{subeqnar} 
\usepackage{multicol} 
\usepackage{cropmark} 
\usepackage{eso}      
\makeindex            

\begin{document}
\title*{Measuring the Mass of the Black Hole in\protect\newline GS2000+25
Using IR Ellipsoidal Variations}
\toctitle{Measuring the Mass of the Black Hole in
\protect\newline GS2000+25 Using Infrared 
\protect\newline Ellipsoidal Variations}
%
%
\titlerunning{Mass of the Black Hole in GS2000+25}
%
\author{Dawn M. Leeber\inst{1}
\and Thomas E. Harrison\inst{1}
\and Bernard J. McNamara\inst{1}}
\authorrunning{Dawn Leeber et al.}
%
%
\institute{New Mexico State University, Las Cruces NM 88003, USA}

\maketitle              

\begin{abstract}
Soft X-ray Transients (SXTs) are binary systems that are believed to consist 
of a black hole and a normal late type dwarf star which fills its Roche Lobe.
We have used GRIM II on the ARC 3.5 meter telescope at Apache Point 
Observatory to obtain infrared photometry of GS2000+25 (QZ Vul).  By modeling 
the SXT ellipsoidal variations with WD98, we can determine the orbital 
period and inclination of the system.  The inclination for a best fit circular 
orbit is $75^o$, and when combined with the observed mass function, corresponds
 to a primary mass of 6.55$M_{\odot}$.  More data is needed to better define 
the minima, fill in the small gaps in the light curve, and explore the 
possibility of an eccentric orbit.  
\end{abstract}

\section{What are Soft X-Ray Transients?}

Soft X-ray Transients\index{Soft X-ray Transients} (SXTs) are {\it transient} 
binary systems that consist of a black hole primary and a 
distorted (i.e. Roche lobe filling), late type, cool dwarf secondary. They 
display large and sudden {\it x-ray} and optical outbursts $(L_x = 10^{38} \ts 
erg/s \ts$; $\Delta V\sim 7 \ts$mags), separated by long intervals of 
quiescence.  Their outbursts result from a sudden, dramatic increase in the 
accretion rate onto the compact object. SXTs show a {\it soft} spectra 
compared to other x-ray sources.  The secondary stars are visible and dominate 
the systemic infrared luminosity in quiescence, allowing us to observe and 
model the ellipsoidal variations of the secondary star as it orbits the 
primary.

\section{IR Ellipsoidal Variations: Observations and Modeling}

The most difficult parameter to determine when estimating the mass of 
the primary is the orbital inclination, $i$.  Since the known SXTs are 
not eclipsing, $i$ can only be determined through the modeling of ellipsoidal 
variations. Ellipsoidal variations result from the rotational and tidal 
distortions of the Roche lobe filling secondary star and its non-uniform 
surface brightness distribution (limb darkening, gravity brightening).  In 
the infrared, the secondary stars of STXs are brighter, and the variations 
are much less contaminated by any residual accretion disk or hot spot than in 
the optical regime.

To construct the GS2000+25\index{GS2000+25} J-band light curve shown in 
Fig. 1, differential photometry was performed on data taken with the 
GRIM II infrared imager on the ARC 3.5 meter telescope at Apache Point 
Observatory. The newest version of the University of Calgary's Wilson-Devinney 
light curve modeling code, WD98\index{WD98}, was used to model the data (Dr. 
Josef Kallrath, private communication).  The model in Fig. 1 includes 
a semi-detached system with a K5V secondary, circular orbit, reflection effect,
 logarithmic limb darkening coefficients, and a J-band Kurucz atmosphere. 

\begin{figure}
\centering
\includegraphics[width=.71\textwidth]{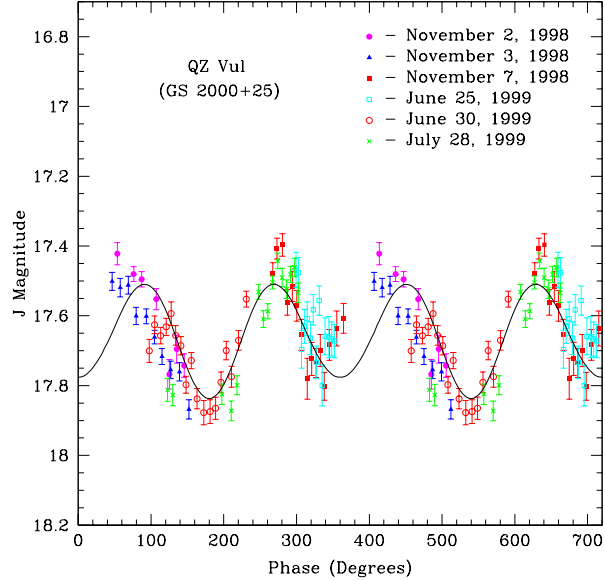}
\caption[]{GS2000+25 J-band light curve and model. Various points represent 
data obtained between 11/98 and 7/99, and the solid line is the $i=75^o$ 
circular orbit model from WD98 described in the text.  Error bars are 
$1\sigma$.  The data show a maximum to minimum amplitude of $\sim 0.4$ mags.}  
\label{eps1}
\end{figure}

\section{Discussion}

The current WD98 model indicates an inclination angle of $75^o$ and a 
resulting mass of $6.55\ts M_{\odot}$.  No outburst x-ray eclipses were 
observed, so $75^o$ represents the upper limit of allowable inclinations. More
data is needed to fill in the gaps in the light curve and better define the 
minima. Each night of data has been carefully normalized, but 
until we have a data set that covers both a minima and maxima, the true 
maximum to minimum amplitude can not be conclusively determined.  Note that 
the circular orbit model does not fit the light curve perfectly.  The 
possibility of an eccentric orbit will be explored once the minima are more 
clearly defined.  If this system were to have an elliptical orbit, it would 
be an unexpected, yet important discovery.

\clearpage
\addcontentsline{toc}{section}{Index}
\flushbottom
\printindex

\end{document}